# Function Overloading Implementation in C++

## A Static Type of Polymorphism

Dr. Brijender Kahanwal
Dept. of Computer Science & Engineering
Galaxy Global Group of Institutions
Ambala, India
imkahanwal@gmail.com

Dr. T. P. Singh
Dept. of Applied Sciences
RPIIT
Karnal, Haryana
tps5675@gmail.com

*Abstract*— In this article the function overloading in object-oriented programming is elaborated and how they are implemented in C++. The language supports a variety of programming styles. Here we are describing the polymorphism and its types in brief. The main stress is given on the function overloading implementation styles in the language. The polymorphic nature of languages has advantages like that we can add new code without requiring changes to the other classes and interfaces (in Java language) are easily implemented. In this technique, the run-time overhead is also introduced in dynamic binding which increases the execution time of the software. But there are no such types of overheads in this static type of polymorphism because everything is resolved at the time of compile time. Polymorphism; Function Overloading; Static Polymorphism; Overloading; Compile Time Polymorphism.

## INTRODUCTION

The word polymorphism is used in coding. It has the origin from Ancient Greek words. Here Poly means many and morph means form. Basically it is the ability to assume several different forms. It is used in Chemistry as the same. In Microbiology (the branch of biology that studies microorganisms and their effects on humans), pleomorphism word is used, it means the ability of some bacteria to alter their shape or size in response to environmental conditions.

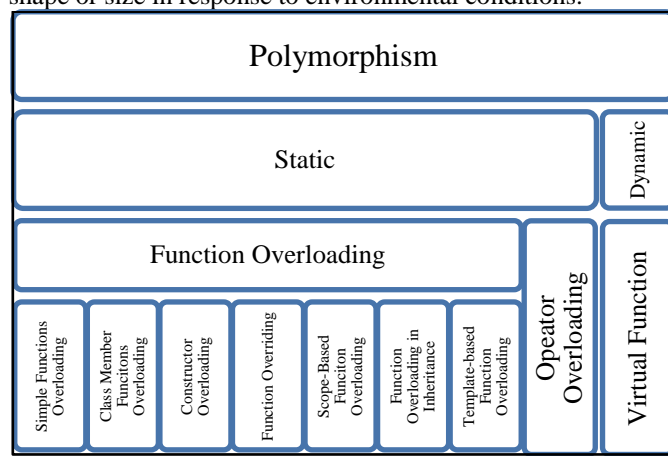

Fig. 1. Taxonomy of function overloading

In Cytology (a branch of Biology concerned with the structure and function of plant and animal cells, pleomorphism means variability in the size and shape of cells and/or their nuclei. The taxonomy of the function overloading is shown in the Fig. 1.

### STATIC POLYMORPHISM

*Function Overloading*

C++ allows specification of more than one function of the same name in the same scope. These are called overloaded functions and are described in detail in Overloading. Overloaded functions enable programmers to supply different semantics for a function, depending on the types and number of arguments. It is explored in the section 4 of this paper.

*Operator Overloading*

It is ad-hoc type of polymorphism. It allows the programmer to redefine how standard operators work with class objects (Class - user-defined data type).We always consider these two things – first is the overloaded operators for classes should behave like operators for built-in types and the overloaded operators should have the same semantics as their built-in counterparts.

### DYNAMIC POLYMORPHISM

The polymorphism exhibited at run time is called dynamic (run-time determination) polymorphism. What makes more sense is to use *run-time binding* where the decision about which version of the method to invoke is delayed until run-time.

In case of static polymorphism, the compiler resolves which function to call at the compile time. But in the case of dynamic polymorphism, this decision is delayed until the runtime.

*Vitual Function*

It is a dynamic polymorphism a virtual function is a member function that you expect to be redefined in derived classes. When you refer to a derived class object using a pointer or a reference to the base class, you can call a virtual function for that object and execute the derived class's version of the function.

Virtual functions ensure that the correct function is called for an object, regardless of the expression used to make the function call. A virtual function is a function in a base class that forms part of the interface for a set of derived classes.

It is declared virtual in the base class and may or may not have a definition in that class. It will have definitions in one or more of the derived classes. The purpose of a virtual function is to have one name, one prototype, and more than one definition so that the function's behavior can be appropriate for each of the derived classes.

## FUNCTION OVERLOADING

In it the same variable name is used to denote different functions, and the context is used to decide which function is denoted by a particular instance of the name. Function (Method) overloading Rules:
 i. All the overloading functions must have the same function name.
 ii. The number of arguments may differ.
 iii. The data types of arguments may differ.
 iv. The order of arguments may differ.
 v. The return type may differ. But only the return type difference is not counted in the function overloading. (see the note)

**Note:** Methods (or functions) with difference in only the return type of prototypes and same in both the number of arguments and as well as the data types of arguments, are treated same by the compiler because the compiler considers only the parameters. The following prototypes are treated same by the compiler.

> int sum (int a, int b);
> float sum (int a, int b);

Function overloading a method (function) allows you to create functions of the same name that take different data types of arguments or different in number of arguments or different in the sequence of arguments as follows:
 a. Methods (or functions) with difference in the data type of arguments of prototypes.
    > int sum (int a, int b);
    > int sum (int a, float b);
 b. Methods (or functions) with difference in the number of arguments of prototypes.
    > int sum (int a, int b);
    > int sum (int a, int b, int c);
 c. Methods (or functions) with difference in the sequence of arguments of prototypes.
    > int sum (int a, float b);
    > int sum (float a, int a);

The function overloading is further explored in the following subsections.

### *Simple Function Overloading*

It is also supported by the C language as in the C++.

```cpp
#include<iostream>
using namespace std;
int sum (int a, int b); // function prototype with two same data type arguments
int sum (int a, float b); // function prototype with two different data type arguments
int sum (float a, int b); // function prototype with two different data type arguments
//with different sequence from the above
int sum (int a, int b, int c); // function prototype with three same data type arguments
int main( )
{
    int x, y, z;
    float p;
    cout<<"Enter the three integers"<< endl;
    cout<<"x=";
    cin>>x;
    cout<<"y=";
    cin>> y;
    cout<< "z=";
    cin>>z;
    cout<< "Enter one float"<<endl<< "p=";
    cin>>p;
    cout<< "Call of int sum (int a, int b);"<< endl;
    cout<< "Result: "<< sum (x, y)<< endl;
    cout<< "Call of int sum (int a, int b, int c);"<<   endl;
    cout<< "Result: "<< sum (x, y, z)<< endl;
    cout<< "Call of int sum (int a, float b);"<< endl;
    cout<< "Result: "<< sum (x, p)<< endl;
    cout<< "Call of int sum (float a, int b);"<< endl;
    cout<< "Result : "<< sum (p, y)<< endl;
    return 0;
}
int sum (int a, int b)
{
    return(a+b);
}
int sum (int a, float b)
{
    return(a+b);
}
int sum (float a, int b)
{
    return(a+b);
}
int sum (int a, int b, int c)
{
    return(a+b+c);
}
```

OUTPUT
Enter the three integers
x=10
y=30
z=50
Enter one float
p=25.5

```
Call of int sum (int a, int b);
Result:  40
Call of int sum (int a, int b, int c);
Result:  90
Call of int sum (int a, float b);
Result:  35
Call of int sum (float a, float b);
Result:  55
Press any key to continue
```

Fig. 2. Simple function overloading and its output

*Class Member Functions Overloading*

```cpp
#include<iostream>
using namespace std;
class calculator
{
public:
        void sum (int x, int y);
        void sum (float x, float y);
        void subtraction (int x, int y);
        void subtraction (float x, float y);
};
void calculator:: sum (int x, int y)
{
        cout<<"Result : "<< (x+y)<< endl;
}
void calculator:: sum (float x, float y)
{
        cout<< "Result : "<< (x+y)<< endl;
}
void calculator:: subtraction (int x, int y)
{
        cout<< "Result : "<< (x-y)<< endl;
}
void calculator:: subtraction (float x, float y)
{
        cout<< "Result : "<< (x-y)<< endl;
}
int main( )
{
        calculator  object;
        int a, b;
        float c, d;
        cout<< "Enter two integers :";
        cin>>a>>b;
        cout<< "Enter two floats :";
        cin>>c>>d;
        cout<< " ----------Two integer's summation : ----------"<< endl;
        object.sum(a, b);
        cout<< endl << " ----------Two float's summation : ----------"<< endl;
        object.sum(c, d);
        cout<< endl << " ----------Two integer's subtraction : ----------"<< endl;
        object.subtraction(a, b);
        cout<< endl << " ----------Two float's  subtraction : ----------"<< endl;
        object.subtraction(c, d);
        return 0;
}
OUTPUT
Enter two integers   :15
25
Enter two floats   : 2.5
5.5
----------------Two integer's summation  : -------------
Result   : 40
----------------Two float's summation  : -------------
Result   : 8
----------------Two integer's subtraction  : -------------
Result   : -10
----------------Two float's subtraction  : -------------
Result   : -3
Press any key to continue
```

Fig. 3. Class member functions overloading and its output

*Constructor (a special type of class member function) Overloading*

Constructors are the special type of member functions of a class. These are called special because of the following properties:
- o  A constructor's name is same as the name of the class, concerned.
- o  These do not have a return value.
- o  These are called automatically when an object of the class is created.
- o  These are always declared in the public section of the class.
- o  These are used to initialize the area fields for the concerned objects.

Other properties are same as of the normal functions or member functions of a class have. So these are also overloaded as the other functions.

```cpp
#include<iostream>
using namespace std;
class A
{
   private:
       int a, b;
   public:
      A( );// default constructor
      A(int x);//constructor with one argument
      A(int x, int y);// constructor with two arguments
      A(A &p); // copy constructo
      void get ( );
      void display ( );
};
   A::A( )
   {
     a=b=0;
   }
   A::A(int x)
   {
```

```
        a=b=x;
    }
    A::A(int x, int y)
    {
       a=x;
       b=y;
    }
    A::A (A &p)
    {
       a=p.a;
       b=p.b;
    }
void A:: get ( )
{
    cin>> a>>b;
}
void A:: display ( )
{
    cout<< "a=  "<< a<<"          "<<"b=  "<< b<< endl;
}
int main( )
{
    A b1, b2(20), b3(250,80), b4(b3);
    cout<< "Use of  default constructor" << endl;
    b1.display( );
    cout<< "Use of  one argumented constructor" << endl;
    b2.display( );
    cout<< "Use of two argumented  constructor" << endl;
    b3.display( );
    cout<< "Use of copy  constructor object b1 is copied in the object b4" << endl;
    b4.display( );
    return 0;
}
OUTPUT
Use of default constructor
a=  0              b=  0
Use of one argumented constructor
a=  20             b=  20
Use of two argumented constructor
a=  250            b=  80
Use of copy constructor object b1 is copied in the object b4
a=  250            b=  80
Press any key to continue
```

Fig. 4. Class constructors overloading and its output

*Function Overriding*

The function overriding takes place in inheritance (an OOPS concept). It allows replacing an inherited method with a different implementation under the same name (same prototype). Usually the overridden function will have the same number, order, and types of arguments. Since, it is proposed to be a matching replacement.

Under this concept both the classes (base class and derived class) have member functions with same name and arguments under the public access specifier and inherited the base class publicly. If we create an object of derived class and write code to access that member function then, the member function in derived class is only invoked. Here, the member function of derived class overrides the member function of base class. It can be resolved with the help of scope resolution operator (::), such as object.base::function_name ( arg1, arg2, …).

```
#include<iostream>
using namespace std;
class A
{
   private:
       int a, b;
    public:
       void get ( );
       void display ( );
};
void A:: get ( )
{
    cin>> a>>b;
}
void A:: display ( )
{
    cout<< "a=  "<< a<<"          "<<"b=  "<< b<< endl;
}
class B : public A
{
   private:
       int c, d, e;
    public:
       void get ( );
       void display ( );
};
void B:: get ( )
{
    cin>> c>>d >>e;
}
void B:: display ( )
{
    cout<< "c=  "<< c<<"          "<<"d=  "<< d<<"  "<< "e=  "<< e << endl;
}
int main( )
{
   B b1;
   cout<< "... Class B's Object's function calls. : Overrided one" << endl;
   cout<< "Enter three integers:" << endl;
   b1.get( );
   cout<< "Entered values are as follows: overrided call" << endl;
   b1.display( );
   cout<< "... Class A's Object's function calls. : resolving of overriding with the help of ::        operator" << endl;
   cout<< "Enter two integers:" << endl;
   b1.A::get( );
   cout<< "Entered values are as follows:" << endl;
```

```
     b1.A::display( );
     return 0;
}
OUTPUT
… Class B's Object's function calls.  : Overrided one
Enter three integers:
45
55
65
Entered values are as follows: overrided call
c=  45             d=  55            e=  65
… Class A's Object's function calls.  : resolving of overriding with the help of
f  :: operator
Enter two integers:
25
35
Entered values are as follows:
a=  25             b=  35
Press any key to continue
```

Fig. 5. Function overriding with its output

*Scope-based Function Overloading*

Some methods are implemented in different scope with the same prototype (Function Header). The scopes cannot overlap with each other. There semantics may be different but the same prototype. The type signature may be different. It is shown in the Fig. 6 and the corresponding implementation is shown in the Fig. 7.

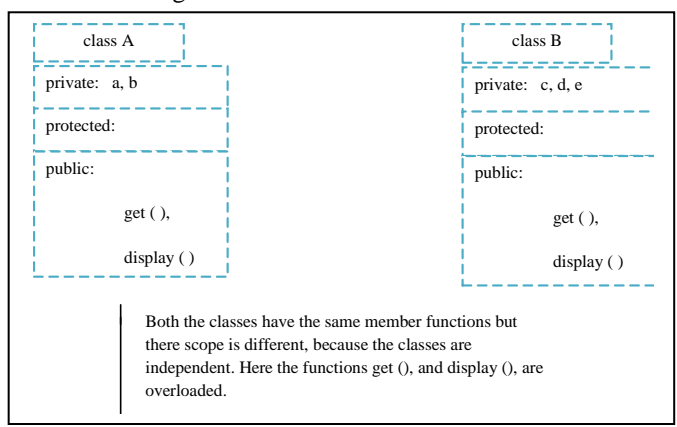

Fig. 6. Scope-based Function overriding of two classes

```
#include<iostream>
using namespace std;
class A
{
   private:
       int a, b;
    public:
       void get ( );
       void display ( );
};
void A:: get ( )
{
     cin>> a>>b;
}
void A:: display ( )
{
     cout<< "a=   "<< a<<"         "<<"b=   "<< b<< endl;
}
class B
{
   private:
       int c, d, e;
    public:
       void get ( );
       void display ( );
};
void B:: get ( )
{
     cin>> c>>d >>e;
}
void B:: display ( )
{
     cout<< "c=   "<< c<<"         "<<"d=   "<< d<<"  "<< "e=   "<< e << endl;
}
int main( )
{
   A a1;
   B b1;
   cout<< "... Class A's Object's function calls. :" << endl;
   cout<< "Enter two integers:" << endl;
   a1.get( );
   cout<< "Entered values are as follows:" << endl;
   a1.display( );
   cout<< "... Class B's Object's function calls. :" << endl;
   cout<< "Enter three integers:" << endl;
   b1.get( );
   cout<< "Entered values are as follows:" << endl;
   b1.display( );
   return 0;
}
OUTPUT
… Class A's Object's function calls.  :
Enter two integers:
10
20
Entered values are as follows:
a=  10             b=  20
… Class B's Object's function calls.  :
Enter three integers:
50
60
70
Entered values are as follows:
c=   50          d=  60           e=   70
Press any key to continue
```

Fig. 7. Scope-based function overloading and its output

*Function Overloading in inheritance*

```cpp
#include<iostream>
using namespace std;
class integer
{
   public:
      void sum (int x, int y);
      void subtraction (int x, int y);
};
void integer:: sum (int x, int y)
{
      cout<< "Result : "<< (x+y)<< endl;
}
void integer:: subtraction (int x, int y)
{
      cout<< "Result : "<< (x-y)<< endl;
}
class real:public integer
{
   public:
      void sum (float x, float y);
      void subtraction (float x, float y);
};
void real:: sum (float x, float y)
{
      cout<< "Result : "<< (x+y)<< endl;
}
void real:: subtraction (float x, float y)
{
      cout<< "Result : "<< (x-y)<< endl;
}
int main( )
{
   real object;
   int a, b;
   float c, d;
   cout<< "Enter two integers :";
   cin>>a>>b;
   cout<< "Enter two floats :";
   cin>>c>>d;
   cout<< " -----------Two integer's summation : ----------"<< endl;
   object.sum(a, b);
   cout<< endl << " -----------Two float's summation : ----------"<< endl;
   object.sum(c, d);
   cout<< endl << " -----------Two integer's subtraction : ----------"<< endl;
   object.subtraction(a, b);
   cout<< endl << " -----------Two float's  subtraction : ----------"<< endl;
   object.subtraction(c, d);
   return 0;
}
OUTPUT
Enter two integers  :20
50
Enter two floats   :2.5
5.9
 -------------Two integer's summation : ------------
Result  :  70
 -------------Two float's summation : ------------
Result  :  8.4
 -------------Two integer's subtraction : ------------
Result  :  -30
 -------------Two float's subtraction : ------------
Result  :  -3.4
Press any key to continue
```

Fig. 8. Function overloading in inheritance and its output

*Template-based Function Overloading*

A template function may be overloaded either by template functions or ordinary functions of its name. In such situations the overloading resolution is accomplished as follows:
 o Call an ordinary function that has an exact match.
 o Call a template function that could be created with an exact match.
 o Try normal overloading resolution to ordinary and call the one that matches

```cpp
#include<iostream>
using namespace std;
template <class S>
void display(S a)
{
    cout<<endl<<"The display template function with one argument :"<< a<<endl;
}
template <class S, class T>
void display(S a, T b)
{
      cout<<endl<<"The display template function with two arguments : "<<a<<"  "<<b<<endl;
}
void display (int a)
{
     cout<< endl<<"Ordinary display function call: "<<a<<endl;
}
int main ( )
{
    display ( 2.6);
    display (2, 4.6);
    display (3.4, 10);
    display ( 'P');
    display (110);
    return 0;
}
OUTPUT
The display template function with one argument  : 2.6
The display template function with two arguments  : 2    4.6
The display template function with two arguments  : 3.4   10
The display template function with one argument :P
```

```
Ordinary display function call : 110
Press any key to continue
```

Fig. 9. Template-based function overloading and its output

## IMPLEMENTATION

All the implementations were done on the compiler Microsoft Visual C++ 6.0. The operating system was Microsoft Windows XP Version 2002 Service Pack 3. The Primary memory (RAM) is of 2 GB. The processor is Intel(R) Core (TM) 2 Duo CPU E7400 @ 2.80GHz. All the programs are also run on the freely available online compiler C++ 4.7.2 (gcc-4.7.2) on ideone.com. It is an online compiler and debugging tool which allows us to compile and run code online in more than 40 programming languages. The web browser, Google Chrome was used to access the website www.ideone.com.

## CONCLUSION & FUTURE SCOPE

This is basically the empirical study of the function overloading, the C++ programming language concept. It is the static polymorphism which may have many variations in the implementation. This paper is good for the students and C++ language lovers to understand the function overloading deeply. In future works we are going to elaborate the dynamic polymorphism, a powerful mechanism in the object oriented programming languages like C++.

## ACKNOWLEDGMENT

The authors are highly grateful to the anonymous reviewers who have commented on this paper. Their detailed and constructive feedback helps a lot in representing the contents and the presentation of the paper.